\documentclass[pra,aps,showpacs,nofootinbib]{revtex4}
\usepackage{epsfig}

\begin{document}

\title{Thermalization in a quasi-1D ultracold bosonic gas}
\author{I. E. Mazets$^{1,2,3}$  and J. Schmiedmayer$^1$}
\affiliation{$^1$Atominstitut, TU Wien, Stadionallee 2, 1020 Vienna, Austria \\
$^2$Ioffe Physico-Technical Institute, 194021 St.Petersburg, Russia,\\
$^3$Wolfgang Pauli Institute, Nordbergstrasse 15, 1090 Vienna, Austria}

\begin{abstract}
We study the collisional processes that can lead to thermalization in one-dimensional systems.  For two body collisions excitations of transverse modes are the prerequisite for energy exchange and thermalzation.  At very low temperatures excitations of transverse modes are exponentially suppressed, thermalization by two body collisions stops and the system should become integrable. In quantum mechanics virtual excitations of higher radial modes are possible. These virtually excited radial modes give rise to effective three-body velocity-changing collisions which lead to thermalization. We show that these three-body elastic interactions are suppressed by pairwise quantum correlations when approaching the strongly correlated regime.  If the relative momentum ${k}$ is small compared to the two-body coupling constant $c$ the three-particle scattering state is suppressed by a factor of $({k}/c)^{12}$, which is proportional to $\gamma ^{12}$, that is to the square of the three-body correlation function at zero distance in the limit of the Lieb-Liniger parameter $\gamma \gg 1$. This demonstrates that in one dimensional quantum systems it is not the freeze-out of two body collisions but the strong quantum correlations which ensures absence of thermalization on experimentally relevant time scales.
\\

\vspace*{1cm}

{\em Short title}: {\bf Thermalization in a quasi-1D ultracold bosonic gas}

\end{abstract}

\maketitle


\section{Introduction} 

One-dimensional (1D) systems \cite{Popov1983,Giamarchi2003} are a model to study the fundamental processes of dynamics and (de)coherence in interacting many-body quantum systems. Ultracold atoms in strongly elongated  traps with $\omega _r \gg \omega_z$ ($\omega _r$, $\omega _z$ being the frequencies of the radial and longitudinal confinement, respectively) offer the possibility to implement  1D quantum physics if both the temperature $T$ and chemical potential $\mu$ are small compared to the energy scale given by the transverse confinement:
\begin{equation}
\mu < \hbar \omega_r, \qquad k_BT<\hbar \omega_r . \label{eq:1}
\end{equation}
1D systems of ultra-cold atoms were implemented in both  optical lattices \cite{OLreview} and atom chips \cite{ACRev}. In the limit of zero temperature they are a realization of the Lieb-Liniger model \cite{LL} of spinless bosons with contact (point-like) interaction,  a prime example of an integrable system.

An important parameter characterizing an 1D system of bosons with point-like interactions described by the (three dimensional) $s$-wave scattering length $\alpha _s$ is the Lieb-Liniger parameter  \cite{LL} 
\begin{equation} 
\gamma =\frac {2\alpha _s}{n_{1D} l_r^2} ,
\label{eq:1a}
\end{equation} 
where $m$ is the mass ot the bosonic atom, $n_{1D}$ the linear density of the atoms in the 1D trap characterized by the transverse confinement frequency $\omega _r$, and  $l_r$ is the fundamental length scale of the localization of an atom in the transversal direction given by  
\begin{equation} 
l_r =\sqrt{\hbar /(m \omega _r)} . 
\label{eq:1b} 
\end{equation} 
The limit $\gamma \ll 1$ corresponds to a weakly-interacting regime, whereas  $\gamma \gg 1$ signifies strongly-interacting, strongly 
correlated (Tonks-Girardeau) regime. 

In an integrable system \cite{th1,yurolw} the number of their integrals of motion equals exactly the number of their degrees of freedom. Thus such a system always ``remembers'' its initial state in the course of its dynamical evolution, and thermalization does not occur. In an integrable system the finite spread of  initial energy may lead only to  relaxation towards the generalized Gibbs (or fully constrained thermodynamic) ensemble \cite{dn1}. Strictly speaking, there is no thermalization in any {\em closed} system, but for non-integrable systems the eigenstate thermalization hypothesis \cite{dn2} holds, enabling  {\em dephasing} to mimic the relaxation to the thermal equilibrium. 

Recently strong inhibition of thermalization, was observed on an optical lattice experiment with bosons deep in the 1D regime \cite{dw1}. On the other hand interference experiments on atom chips with pairs of weakly interacting Bose gases  fulfilling the conditions of Eq. (\ref{eq:1}) are in a good agreement with the thermal-equilibrium description of the 1D atomic ensembles \cite{js1,va,js2}. 

In the present paper we investigate the collisional properties of Bose gases in a 1D geometry and how they contribute to thermalization.    
We follow thereby a procedure outlined in our two previous works \cite{we1,we2}, and give a more in-depth description of the underlying theoretical considerations.  We start with the calcualtion of the freeze-out of thermalization providing two body collisions.  We then proceed to show that virtual excitations of excited states lead to effective three-body collisions, which lead to a term in the Hamiltonian  that breaks integrability and enables thermalization. We then proced to estimte the effects of quantum correlations in 1D and show how they suppress the three-body term for strongly correlated 1D systems, thereby extending the time scale, on which a quasi-1D system can be \textit{approximately} described as integrable. 

In our theoretical considerations we consider identical bosons in a tight 1D wave guide with radial confinement given by a 2D harmonic oscillator with frequency $\omega _r$ (we set $\omega _z=0$).  The contact interaction is described by the pseudopotential $4\pi \hbar ^2m^{-1}\alpha _s \delta ({\bf r}-{\bf r}^\prime )$. The Hamiltonian of the 1D system is:
\begin{eqnarray}
    \hat{\cal H}_{3D}&=&\int d^3{\bf r}\,
    \left[ \hat{\psi }^\dag ({\bf r})
    \left( -\frac {\hbar ^2}{2m}\frac {\partial ^2}{\partial z^2}
    +\hat{H}^{(r)}  \right)
    \hat{\psi } ({\bf r})+ \right. \nonumber \\
    && \left.  \frac {2\pi \hbar ^2\alpha _s}m \hat{\psi }^\dag ({\bf r})
    \hat{\psi }^\dag ({\bf r}) \hat{\psi }({\bf r})\hat{\psi }({\bf r})
    \right] , \label{eq:2} \\
    \hat{H}^{(r)}&=&-\frac { \hbar ^2}{2m}\left(
    \frac {\partial ^2}{\partial x^2}+\frac {\partial ^2}{\partial y^2}
    \right) +\frac{m\omega _r^2}2 (x^2+y^2).    \label{eq:3}
\end{eqnarray}
Thereby the field operators $\hat{\psi }({\bf r})$ are assumed to vanish for $x^2+y^2\rightarrow \infty $ 
and to be periodic along $z$ with the period $L$. For the solutions to Eq. \ref{eq:3} we make the \textit{Ansatz}:
\begin{eqnarray}
    \hat{\psi }({\bf r})&=&\sum_{n,\ell ,k}\hat{a}_{ \{n,\ell \}\, k} \varphi_{n,\ell }(x,y)\frac {\exp (ikz)}{\sqrt{L}}.
    \label{eq:4}
\end{eqnarray}
where $L$ is the quantization length and the atomic annihilation and creation operators $\hat{a}_{ \{n,\ell \}\, k} $ and $\hat{a}_{ \{n,\ell \}\, k}^\dag $ obey the standard bosonic commutation rules.  $\varphi _{n,\ell }(x,y)$ is the normalized eigenfunction of both the radial confinement Hamiltonian, 
$$\hat{H}^{(r)}\varphi _{n,\ell }(x,y)=(n+1)\hbar \omega _r \varphi _{n,\ell }(x,y)$$ 
and the $z$-projection of the orbital momentum, 
$$-i[x(\partial /\partial y)-y(\partial /\partial x)] \varphi _{n,\ell }(x,y)=\ell  \varphi _{n,\ell }(x,y).$$ 
Because we consider identical Bosons the main quantum number $n=0,\, 1,\, 2,\, \dots \, $, and the $z$-projection quantum number $\ell $ of the orbital-momentum is restricted by 
$$|\ell |= \mathrm{mod}\, (n,2),\, \mathrm{mod}\, (n,2)+2,\,\dots \, ,\, n-2,\, n$$ 
and thus has the same parity as the main quantum number. 

\section{Two-Body collisions} 

We first look at collisions of two identical bosonic atoms that are initially in the transverse ground state of the radial confinement.  If the collision is restricted to 1D, that is both atoms remain after the collision in the  transverse ground state, then there can be no energy exchange and cinsequently no thermalization. For such two-body collisions to contribute to energy exchange and thermalization, they have to lead to a change in transverse excitation. By symmetry $\Delta n_1 + \Delta n_2$ must be even. For atoms in the transverse ground state ($n_1=n_2=0$) $\Delta n_1 = \Delta n_2$ band following the above considerations their orbital-momentum quantum numbers after collision are restricted to $-\ell $ and $+\ell $. The rate of populating the radially excited modes by pairwise atomic collisions $\Gamma _{2b}$, can then be estimated for a non-degenerate Bose gas, using Fermi's golden rule. For $k_BT<\hbar \omega _r$ this rate is
\begin{eqnarray}
    \Gamma _{2b} \approx \frac{2\sqrt{2} \hbar n_{1D}\alpha_s^2}{ml_r^3} \; e^{-\frac {2\hbar \omega _r}{k_BT}} =
    2\sqrt{2} \omega _r \, \zeta \, e^{-\frac {2\hbar \omega _r}{k_BT}}. 
    \label{eq:20}
\end{eqnarray}
The dimensionless quantity 
$$\zeta =n_{1D}\alpha _s^2/l_r$$ 
combines two dimensionless parameters characterizing a 1D system.  $n_{1D}\alpha _s \propto \frac{\mu}{\hbar \omega_r}$ is a measure how much the interaction energy (the chemical potential $\mu$) is smaller than the energy scale given by the transverse confinement.  $\alpha _s/l_r$ characterizes the relation between the transverse confinement and the strength of the contact interaction. Its importance can be seen when looking at how the effective 1D coupling constant $g_{1D}$ of pairwise interacting bosonic atoms in a waveguide changes with confinement due to virtual excitation of the radial modes.  Following Olshanii \cite{ol1} $g_{1D}=2\hbar \omega _r \alpha _s /[1-\frac{C^\prime \alpha _s}{\sqrt{2}l_r}]$, $C^\prime \approx 1.46$ and increases as $\alpha _s/l_r$ grows. This points to the ratio $\alpha  _s/l_r\ll 1$ as the measure how much the 1D approximation is violated. In a general case we have to change in all the following expressions $\alpha _s$ to $ \alpha _s /[1-\frac{C^\prime \alpha _s}{\sqrt{2}l_r}]$.

Eq. (\ref{eq:20}) has a also transparent physical interpretation:  The rate $\Gamma _{2b}$ is related to the 3D atomic density ($\sim n_{1D}/l_r^2$), times the $s$-wave scattering cross-section ($\sim \alpha _s^2$), times the exponential Boltzmann factor for the fraction of atoms fast enough to scatter into higher radial modes, times the corresponding velocity of the collision ($\sim \hbar
/(ml_r)$).

Looking at the scaling of Eq. (\ref{eq:20}) one immedeately sees that the collision rate leading to thermalization ($\Gamma _{2b}$) rapidly diminishes when the temperature approaches $T \sim \hbar \omega_r$ and is suppressed by more then a factor of 50 ($e^{-4}$) for $T=\frac{1}{2} \hbar \omega_r$. Estimating the numbers for recent experients \cite{js2} in $^{87}$Rb:  $\alpha _s = 5.3~\textrm{nm}$, $n_{1D}=50~ \mu $m$^{-1}$, $\omega_r /(2 \pi )= 3$ kHz, $T=30$ nK ($\zeta \approx 0.007$) one obtains a collision rate of $\Gamma _{2b} \sim 0.02 \, s^{-1}$. Ths is at least one order of magnitude too small for two body collisions to be responsible for the thermalization required in the evaporative cooling process leading to these low temperatures. 

\section{Three-Body collisions} 

If the kinetic energy of the collision is less than $2\hbar \omega_r$, then the radial modes can be excited only virtually.  Such processes contribute to the system dynamics in the second and higher orders of perturbation theory. 

\begin{figure}[t]
 \epsfig{file=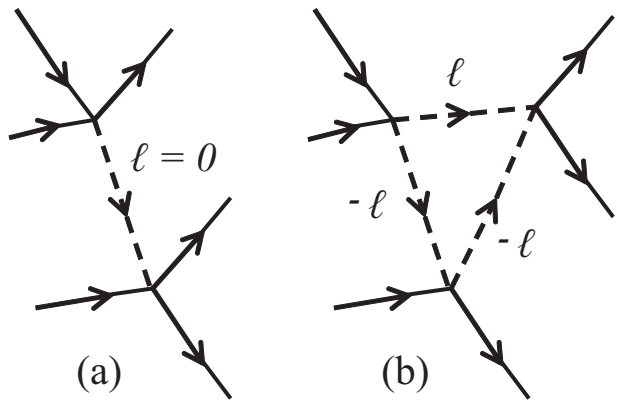,width=0.45\columnwidth}
 \caption {Feynman diagrams for the effective three-body processes in the second \textit{(a)} and third \textit{(b)} orders of perturbation theory. Solid and dashed  lines correspond to atoms in the ground and excited states of the  radial trapping Hamiltonian, respectively.}
 \label{fig:1}
\end{figure}

The simplest case is when after the collision the radial motion state is $| \{ n_1^\prime ,\ell _1^\prime \} , \, \{ n_2^\prime , \ell _2^\prime \} \rangle =| \{ 0,0 \} , \, \{ 2p ,0 \} \rangle $.  Then only one more collision is enough to de-excite the radial mode and bring the system back on the energy shell [see Fig. \ref{fig:1}(a)]. Such a process yields an effective three-body collision already in the {\em second} order of perturbation theory.

In contrast processes involving a virtual excitation to $| \{n_1^\prime ,-\ell \} , \, \{ n_2^\prime ,+\ell \} \rangle $, $\ell \neq 0$, shown in Fig.~\ref{fig:1}(b), contribute only in the {\em third} order, and thus will be neglected.

\subsection{Perturbative approach} 

We will now calculate the matrix elements for the process shown in figure \ref{fig:1}(a) which leads to effective three-body collisions.  In our perturbation calculation the small parameter is $n_{1D}\alpha _s$, that is the mean field interaction $\mu$ is much smaller then the energy scale $\hbar \omega_r$ connected to the transverse confinement.  In addition, to avoid complications related to the confinement-induced resonance in 1D scattering \cite{ol1} we assume $\alpha _s\ll l_r$. We can then rewrite the Hamiltonian (\ref{eq:2}) as
\begin{eqnarray} 
\hat{\cal H}_{3D}&=&\sum_{n,\ell ,k}\left( \frac {\hbar ^2k^2}{2m} +n\hbar \omega _r \right) 
\hat{a}_{ \{n,\ell \}\, k} ^\dag \hat{a}_{ \{n,\ell \}\, k} + \nonumber \\ && 
\frac  {2\pi \hbar ^2\alpha _sf_{0, 0;0, 0}^{0, 0;0, 0}}{m L} \sum _{k, k^\prime ,q} 
\hat{a}_{ \{0,0 \}\, k-q}^\dag \hat{a}_{ \{0,0 \}\, k^\prime +q}^\dag 
\hat{a}_{ \{0,0 \}\, k^\prime }\hat{a}_{ \{0,0 \}\, k} + \nonumber \\ &&
\frac  {4\pi \hbar ^2\alpha _s }{m L} \sum _{k, k^\prime ,q}
\sum _{p=1}^\infty \left( f^{2p, 0;0, 0}_{0, 0;0, 0} 
\hat{a}_{ \{2p,0 \}\, k-q}^\dag \hat{a}_{ \{0,0 \}\, k^\prime +q}^\dag 
\hat{a}_{ \{0,0 \}\, k^\prime }\hat{a}_{ \{0,0 \}\, k} + \mathrm{H.c.}\right) +\hat{\cal R}, 
\label{pert1} 
\end{eqnarray}
where all the terms irrelevant to the process under discussion (Fig.~1a) are gathered in $\hat{\cal R}$, and  the radial matrix element $f^{2p, 0;0, 0}_{0, 0;0, 0}$ is given by: 
\begin{eqnarray}
f^{2p, 0;0, 0}_{0, 0;0, 0} &=& \int dx\int dy \int dx^\prime \int dy^\prime \, 
   \varphi _{n=2p,\ell =0}^*(x,y)\varphi _{0,0}^*(x^\prime ,y^\prime )\delta (x-x^\prime )
\delta(y-y^\prime )\times \nonumber \\ &&  \varphi _{0,0}(x^\prime ,y^\prime )\varphi _{0,0}(x,y), 
\qquad p=0,\, 1,\, 2,\, 3,\, \dots \,  
    \label{eq:5}
\end{eqnarray}
It connects to two atoms in the ground state of the incoming channel, to one atom remaining in the same state, and the other being excited to a state with zero orbital-momentum quantum number and even main quantum number $n=2p$, $p=0,\, 1,\, 2,\, ... $ (remember for Boson $n$ and $\ell $ are required to have the same parity). 

To evaluate Eq. (\ref{eq:5}), we recall that the normalized radial wave functions $\varphi _{n,\ell }(x,y)$ of interest are real and can be expressed through Laguerre polynomials $L_p$
\begin{equation} 
 \varphi _{n=2p,\ell =0}(x,y)= (\pi l_r^2)^{-1}\exp \left( -\frac {x^2+y^2}{2l_r^2}\right) L_p \left( \frac {\sqrt{x^2+y^2}}{l_r}\right) . 
\label{pert2}
\end{equation} 
Since $L_p(0)=1$, we obtain  \cite{yur1o} 
\begin{eqnarray}
\int dx\int dy \int dx^\prime \int dy^\prime \, 
   \varphi _{n=2p,\ell =0}\left( \frac {x-x^\prime }{\sqrt{2}},\frac {y-y^\prime }{\sqrt{2}}\right) \varphi _{0,0}
 \left( \frac {x+x^\prime }{\sqrt{2}},\frac {y+y^\prime }{\sqrt{2}}\right) \times && \nonumber \\  
\delta (x-x^\prime )\delta(y-y^\prime ) \varphi _{0,0}(x^\prime ,y^\prime )\varphi _{0,0}(x,y)&=&\frac 1{2\pi l_r^2}, 
\label{pert3}
\end{eqnarray} 
independently of $p$. Then we easily obtain the necessary matrix element as $f^{2p, 0;0, 0}_{0, 0;0, 0}={\cal C}^{2p,0;0,0}_{2p,0;0,0}/{2\pi l_r^2}$, where the coefficient ${\cal C}^{2p,0;0,0}_{2p,0;0,0}$ is defined by the expansion 
\begin{equation} 
\varphi _{2p,0}\left( \frac {x-x^\prime }{\sqrt{2}},\frac {y-y^\prime }{\sqrt{2}}\right) \varphi _{0,0}
 \left( \frac {x+x^\prime }{\sqrt{2}},\frac {y+y^\prime }{\sqrt{2}}\right)=\sum _n\sum _\ell 
{\cal C}^{2p,0;0,0}_{2p-n,\ell ;n,-\ell }\varphi _{2p-n,\ell }(x,y)\varphi _{n,-\ell } (x^\prime ,y^\prime ). 
\label{pert4} 
\end{equation} 
Comparing the coefficients in front of $(x^2+y^2)^p$ in the left- and right-hand sides of Eq. (\ref{pert4}), we obtain ${\cal C}^{2p,0;0,0}_{2p,0;0,0}=2^{-p}$ and 
\begin{equation} 
f^{2p, 0;0, 0}_{0, 0;0, 0}=\frac 1{2^{p+1}\pi l_r^2 }.
\label{pert5} 
\end{equation}  

In our consideration we are only interested in the case where the collision energy of the two atoms is always much smaller than $\hbar \omega _r$. Then, using the matrix element  (\ref{pert5})  and adiabatically eliminating the radially excited mode operators, we obtain from the original Hamiltonian Eq. (\ref{eq:2}) an effective 1D Hamiltonian:
\begin{equation}
\hat{\cal H}_{1D}=\sum _k \frac {\hbar ^2k^2}{2m} \hat{a}_k ^\dag \hat{a}_k  +\frac {\hbar \omega _r\alpha _s}L
\sum _{k,k^\prime ,q}\hat{a}_{k+q} ^\dag \hat{a}_{k^\prime -q}^\dag \hat{a}_{k^\prime } \hat{a}_k - 
    \frac {\xi \hbar \omega _r\alpha _s^2}{2L^2}
    \sum _{\{ k^\prime _j\} }\hat{a}_{k_1^\prime }^\dag \hat{a}_{k_2^\prime }^\dag
    \hat{a}_{k_3^\prime }^\dag \hat{a}_{k_1}\hat{a}_{k_2}\hat{a}_{k_3}, \label{eq:10}
\label{eq:11a}
\end{equation}
where we write for simplicity $\hat{a}_k \equiv \hat{a}_{ \{0,0\} \, k}$ and the numerical constant $\xi$ is given by 
\begin{equation}
\xi =4 \sum _{p=1}^\infty 1/(4^p p)=4 \, \mathrm{ln}\, (4/3) \approx 1.15 .
\label{defxi}
\end{equation}  
Note that the relative contribution of the virtual states with the excitation energy higher than $2\hbar \omega _r$ given by $(\xi -1)/ \xi $ is remarkably small. The summation in the last term of Eq. (\ref{eq:11a}) is taken over all the kinetic momenta obeying the conservation law 
$$k_1 ^\prime + k_2^\prime + k_3 ^\prime =k_1 + k_2 +k_3 \, .$$  
Introducing the field operator $\hat{\tilde{\psi }}(z)=L^{-1/2}\sum _k \hat{a}_k \exp(ikz) $, we can rewrite Eq. (\ref{eq:11a}) as 
\begin{equation} 
\hat{\cal H}_{1D}=\int dz\, \left( \frac {\hbar^2}{2m} \frac {\partial \hat{\tilde{\psi }}^\dag }{\partial z} 
\frac {\partial \hat{\tilde{\psi }}}{\partial z} +\hbar \omega _r\alpha _s \hat{\tilde{\psi }}^\dag 
\hat{\tilde{\psi }}^\dag \hat{\tilde{\psi }} \hat{\tilde{\psi }}-\frac \xi 2\hbar \omega _r\alpha _s^2
\hat{\tilde{\psi }}^\dag \hat{\tilde{\psi }}^\dag \hat{\tilde{\psi }}^\dag 
\hat{\tilde{\psi }} \hat{\tilde{\psi }} \hat{\tilde{\psi }}\right) . 
\label{eq:11b} 
\end{equation} 
The first and second terms in Eqs. (\ref{eq:11a}) or (\ref{eq:11b}) correspond to the Lieb-Liniger model.  The third (\textit{cubic}) term stems from the effective three-body collisions mediated by virtually excited states \footnote{Note that in our previous work \cite{we1} the coefficient in front of this cubic term was estimated by a factor 4 too large.}. This third (\textit{cubic}) term in Eq. \ref{eq:11b} violates the integrability in the 1D system. 

The fact that the effective three-body interactions are dominated by virtual excitations of the lowest even-parity excited state may seem 
surprising, since the correct calculation of the effective {\em two-body} 1D coupling constant requires taking into account of the 
infinite number of states \cite{ol1}. However, in the latter case one deals with the calculation of the two-body wave function, which has in 1D 
a $1/z$ singularity (stemming from the 3D boundary condition at $r\rightarrow 0$ that provides the correct asymptotic form of the 
scattered $s$-wave). The removal of this singlularity  yields the regular part of the two-body wave function and, through this 
regular part, the scattering amplitude and, hence, the effective coupling in 1D. On the other hand, if one tries to obtain the effective 1D 
interaction by adiabatic elimination of {\em all} the excited states, one gets a divergent series $\sum _{n=0}^\infty n^{-1/2}$ in the 
expression for the effective 1D coupling constant. The aforementioned regularization of the wave functions formally corresponds to the 
renormalization of this divergent series via substituting it by a finite expression $\lim _{s\rightarrow \infty } \left( \sum _{n=0}^s n^{-1/2}-
\int _0^s d\nu \, \nu ^{-1/2} \right) $ \cite{ol1}. In our case, processes related to {\em three-body} collisions do not give rise to additional 
singularities in the many-body wave function, and no additional regularization is needed. The sum over all excited states thus converges. The 
convergence is rapid enough to ensure fair estimation of the whole sum by its first term. 

Accurate calculation of the effective three-body interaction potential ${\cal U}_\mathrm{3b}$ yields 
\begin{equation} 
{\cal U}_\mathrm{3b}(z_1,z_2,z_3) =-\frac 12\hbar \omega _r\alpha _s^2[ {\cal Y}(z_1,z_2;z_3)+{\cal Y}(z_1,z_3;z_2)+{\cal Y}(z_2,z_3;z_1)],
\label{U3bIIIA} 
\end{equation} 
where 
\begin{equation}
{\cal Y}(z_1,z_2;z_3)=[\delta (z_3-z_1)+\delta (z_3-z_2)]\sum _{p=1}^\infty \frac {\sqrt{p}}{\sqrt{2}\, 4^{p-1} p\, l_r}\exp \left( -
\frac{\sqrt{2 p} \, |x_1-x_2|}{l_r }\right) . 
\label{YIIIA}
\end{equation} 
Obviously, the sum in Eq. (\ref{YIIIA}) converges and gives a sharp-peaked function rapidly (exponentially) decreasing at distances 
much larger than $l_r$. Since Eq. (\ref{eq:1}) holds, we consider scattering events with transferred momenta much less than $\hbar /l_r$. In this case we can use approximation ${\cal Y}(z_1,z_2;z_3)\approx \xi [\delta (z_3-z_1)+\delta (z_3-z_2)]\delta (z_1-z_2)$. Then, by taking the 
matrix element of the effective interaction ${\cal U}_\mathrm{3b}$ and dividing it by 3! (the number of permutation of three identical 
particles) we obtain the last term in the second-quantized Hamiltonian (\ref{eq:10}). 

Before continuing we want to point out similarities with other recent works: (1) The mechanism discussed here is to a certain extent similar to the virtual association of atoms to a molecular dimer \cite{yurbro}. In our discussion here, virtual excitation of radial modes during a two-atom collision temporarily localize the interatomic distance on the length scale $\sim l_r$.  Scattering a third atom on such a transient structure of finite size and  mass $2m$ leads to an effective threebody collision. In Ref. \cite{yurbro}, collisions of a third atom bring ``virtual'' dimers, enhanced in size by a Feshbach resonance down to the energy shell, thus bringing about ``quantum chemistry'' in 1D. (2) In a similar way effective three-body interactions between polar molecules emerge, due to virtual transitions to an off-resonant internal state \cite{z3b}.

\subsection{Variational approach} 

The cubic term in Eq. (\ref{eq:11b}) is negative and thus supports no bound ground state. Therefore we have to consider it as a first correction term to the purely pairwise interaction energy in the effective 1D Hamiltonian. Considering the mean-field limit of Eq. (\ref{eq:11b}), $\hat{\tilde{\psi }}\approx \sqrt{n_{1D}}\exp(i\theta)$, we get the energy density (per unit length) 
\begin{equation} 
{\cal E}_{pert}= \frac {\hbar^2}{2m} \left[ \left(\frac {\partial \sqrt{n_\mathrm{1D}} }{\partial z} \right)^2+
n_\mathrm{1D}\left(\frac {\partial \theta }{\partial z} \right)^2\right] +\hbar \omega _r \alpha _s n_{1D}^2 -
\frac \xi 2 \hbar \omega _r \alpha _s^2 n_{1D}^3, 
\label{eq:var1}
\end{equation} 
this expansion is correct in the limit of the small linear density $n_{1D}a_s\ll 1$. If we neglect the contribution of the radial levels with the main quantum number larger than 2 by setting $\xi \approx 1$, we see that Eq. (\ref{eq:var1}) is the expansion up to the cubic term of the energy density obtained by the variational method by Salasnich, Parola and Reatto \cite{Sal2002}
\begin{eqnarray} 
{\cal E}_{var}&=&\frac {\hbar^2}{2m} \left[ \left(\frac {\partial \sqrt{n_\mathrm{1D}} }{\partial z} \right)^2+
n_\mathrm{1D}\left(\frac {\partial \theta }{\partial z} \right)^2\right]+n_{1D}\left( \frac {\hbar ^2}{2m\sigma ^2}+ 
\frac 12 m\omega _r^2 \sigma ^2\right) +\frac {\hbar ^2\alpha _sn_{1D}^2 }{m\sigma ^2}\nonumber \\
&=&\frac {\hbar^2}{2m} \left[ \left(\frac {\partial \sqrt{n_\mathrm{1D}} }{\partial z} \right)^2+
n_\mathrm{1D}\left(\frac {\partial \theta }{\partial z} \right)^2\right] +n_{1D}\hbar \omega _r 
\sqrt{1 + 2n_{1D} \alpha _s} . 
\label{eq:var2}
\end{eqnarray} 
Here one assumes the wave function of the transversal atomic motion to be $\propto \exp [-(x^2+y^2)/(2\sigma ^2)]$, with $\sigma ^2=\frac \hbar {m\omega _r}\sqrt{1 + 2n_{1D} \alpha _s} $ which minimizes ${\cal E}_{var}$. 

\section{Calculations of the collision rates}

We now turn to the collision rates for these effective three body collisions.  We start with calculating the rate $\Gamma _{k_1k_2k_3 }$ for the decay of a specific state $|k_1,k_2,k_3\rangle \equiv a^\dag _{k_1}a^\dag _{k_2}a^\dag _{k_3} |\mathrm{vac}\rangle $, $|\mathrm{vac}\rangle $ being the vacuum state of the atomic field wherein atoms are absent (should not be confused with the {\em vacuum of elementary excitations}) due to three-body collisions. The final states of the decay are $|k_1^\prime ,k_2^\prime ,k_3^\prime \rangle 
\equiv a^\dag _{k_1^\prime }a^\dag _{k_2^\prime }a^\dag _{k_3^\prime } |\mathrm{vac}\rangle $.  To make the calcualtion simple we assume the 1D bosonic gas being strongly non-degenerate ($k_BT$ much higher than the chemical potential) and weakly-interacting. This enables us to neglect the probability double (and higher) occupation of any $k$-mode, therefore assuming all involved atomic momenta to be different, and assume the elementary excitations coinciding with the atomic plane waves with the free-particle (quadratic) dispersion law. Then Fermi's golden rule yields 
\begin{eqnarray} 
\Gamma _{k_1k_2k_3 }&=&\frac {2\pi }\hbar L^2\int \frac {dk_2^\prime }{2\pi }
\int \frac {dk_3^\prime }{2\pi }\,\delta \left[ \frac {\hbar ^2}{2m} \left( \sum _j^3k^{\prime \, 2}_j -\sum _j^3k^{2}_j 
\right) \right] \left( \frac {\xi \hbar \omega _r\alpha _s^2}{2L^2}\right) ^2 \times \nonumber \\ && 
\left. | \langle k_1^\prime ,k_2^\prime ,k_3^\prime | 
\sum _{ \{q_j^\prime \} }  \hat{a}_{q_1}^\dag \hat{a}_{q_2 }^\dag
    \hat{a}_{q_3}^\dag \hat{a}_{q_1}\hat{a}_{q_2}\hat{a}_{q_3}|k_1,k_2,k_3\rangle | ^2
    \right| _{k_1^\prime +k_2^\prime +k_3^\prime =k_1+k_2+k_3} .
\label{eq:rates1} 
\end{eqnarray} 
To account for the condition $k_1^\prime +k_2^\prime +k_3^\prime =k_1+k_2+k_3$ we add an additional integration over $k^\prime _1$ with the necessary delta-function: 
\begin{eqnarray} 
\Gamma _{k_1k_2k_3 }&=&\frac {L^2}{2\pi \hbar }\int \int \int _{{\cal W}^\prime } dk_1^\prime 
{dk_2^\prime }{dk_3^\prime }\,\delta \left[ \frac {\hbar ^2}{2m} \left( \sum _j^3k^{\prime \, 2}_j -\sum _j^3k^{2}_j 
\right) \right] \delta \left( \sum _j^3k^{\prime }_j -\sum _j^3k_j \right) \times \nonumber \\ && 
\left( \frac {\xi \hbar \omega _r\alpha _s^2}{2L^2}\right) ^2 \left| \langle k_1^\prime ,k_2^\prime ,k_3^\prime | 
\sum _{ \{q_j^\prime \} } \hat{a}_{q_1}^\dag \hat{a}_{q_2 }^\dag
    \hat{a}_{q_3}^\dag \hat{a}_{q_1}\hat{a}_{q_2}\hat{a}_{q_3}|k_1,k_2,k_3\rangle \right| ^2  .
\label{eq:rates2} 
\end{eqnarray} 
Since the atoms are indistinguishable, we need to integrate over the volume ${\cal W}^\prime $ in the $k^\prime $-space that corresponds to a unique ordering of the variables, e.g., $k_1^\prime > k_2^\prime > k_3^\prime $. The integral over ${\cal W}^\prime $ of a function fully symmetric over permutations of $k_1^\prime , k_2^\prime , k_3^\prime $ amounts to
$1/3!$ of the integral over the whole $k^\prime $-space. If all the involved momenta are different (as is the case for a system far from degeneracy) all possible ways of ordering  three bosonic creation operators with  lower indices $k_1^\prime ,k_2^\prime ,k_3^\prime $ and three bosonic annihilation operators with  lower indices  $k_1,k_2,k_3$ finally  yield
\begin{equation} 
\langle k_1^\prime ,k_2^\prime ,k_3^\prime | 
\sum _{ \{q_j^\prime \} } \hat{a}_{q_1}^\dag \hat{a}_{q_2 }^\dag
    \hat{a}_{q_3}^\dag \hat{a}_{q_1}\hat{a}_{q_2}\hat{a}_{q_3}|k_1,k_2,k_3\rangle =(3!)^2 
\label{eq:rates3} 
\end{equation} 
To evaluate the integral in Eq. (\ref{eq:rates3}), we perform an orthogonal transformation from $k_1^\prime , k_2^\prime , k_3^\prime $ to Jacobi co-ordinates (here we deal with the 1D analog of the hyperspherical co-ordinates, which are used in the three-body problem \cite{o1}) in the wavenumber space: 
\begin{equation}
k_c^\prime =\frac 1{\sqrt{3}}(k_1^\prime + k_2^\prime + k_3^\prime ),\quad 
k_{12}^\prime =\frac 1{\sqrt{2}}(k_1^\prime - k_2^\prime  ), \quad 
k_{321}^\prime =\sqrt{\frac 23}\left( k_3^\prime -\frac{ k_1^\prime + k_2^\prime }2\right) .
\label{eq:rates4} 
\end{equation} 
Further we introduce the hyperangle $\chi ^\prime $ via 
\begin{equation} 
k_{12}^\prime =\tilde{k}^\prime \sin \chi ^\prime , \qquad 
k_{321}^\prime =\tilde{k}^\prime \cos \chi ^\prime , 
\label{eq:rates4bis} 
\end{equation} 
and express Eq. (\ref{eq:rates3}) as 
\begin{eqnarray}
\Gamma _{k_1k_2k_3 }&=&\frac {(3!)^3\xi ^2m\omega _r^2\alpha _s^4}{4\pi \hbar L^2} 
\int _{-\infty }^\infty dk^\prime _c \int _{-\pi }^\pi d\chi ^\prime 
\int _0^\infty d\tilde{k}^\prime  \, \tilde{k}^\prime \times \nonumber \\ && 
\delta (\sqrt{3}k^\prime _c-k_1-k_2-k_3)\delta (k^{\prime \, 2}_c+\tilde{k}^{\prime  \, 2}-
k_1^2 -k_2^2 -k_3^2 )=3!C_{3b}\frac {\omega _r \alpha _s^4}{L^2l_r^2}, 
\label{eq:rates5} 
\end{eqnarray}
and accurate evaluation of the integral yields (cf. \cite{we1}) 
\begin{equation} 
C_{3b}=3\sqrt{3}\xi ^2\approx 6.88. 
\label{eq:rates5bis}
\end{equation}

To calculate the three-body  collision rate $\Gamma _{3b}$ per atom, we need to multiply $\Gamma _{k_1k_2k_3 }$ by the product of populations $Nf_{k_j}$ of the states $|k_j\rangle = a_{k_j}^\dag |\mathrm{vac}\rangle $, $j=1,2,3$ (the occupation probabilities are normalized to unity, $\int _{-\infty }^\infty dk\, f_k=1$) and integrate the whole $k_1,k_2,k_3$-space, divide by 3! to take into account the indistinguishability of the bosons. Then we obtain the number of three-body collisions per unit time in the whole system. This number should be divided by $N$ to obtain the rate per atom:
\begin{equation}
    \Gamma_{3b}=C_{3b}\frac {\hbar}{m} \left( \frac {n_{1D}}{l_r ^2} \right)^2 \alpha _s ^4
    = C_{3b}  \, \zeta^2 \, \omega_r .
    \label{eq:17b}
\end{equation}
The result of Eq. (\ref{eq:17b}) may seem counterintuitive at first: the collision rate is independent of temperature, and it is proportional to the dimensional parameter $\zeta^2$ and the radial confinement $\omega_r$.

The physics behind the first observation is related to the fact that the collision kinetic energy is small compared to the virtual excitation energy.  This was one of our assumptions in deriving the effective three body collisions and is required by the condition ($k_BT<\hbar \omega_r$) to be fulfilled when building a 1D system (Eq.~(\ref{eq:1})). Consequently the composite matrix element of the second-order process should not depend in leading order on the velocities (energies) of the colliding particles and hence on temperature (see Eq.~(\ref{eq:5})). In addition the phase space volume for the scattered particles is independent on the incoming momenta $k_1$, $k_2$, and $k_3$.

The other terms can be motivated the following basic physics considerations: Since effective three-body elastic scattering is the dominant process the scattering rate must be proportional to the 3D density squared: $(n_{1D}/l^2_r)^2$.  Furthermore, the scattering rate contains the square of the matrix element corresponding to the diagram in Fig.~1(a), where each vertex is proportional to $\alpha_s$, therefore this rate is proportional to $\alpha _s^4$. The factor $\hbar/m$ provides the correct dimensionality (s$^{-1}$).

\begin{figure}
  \epsfig{file=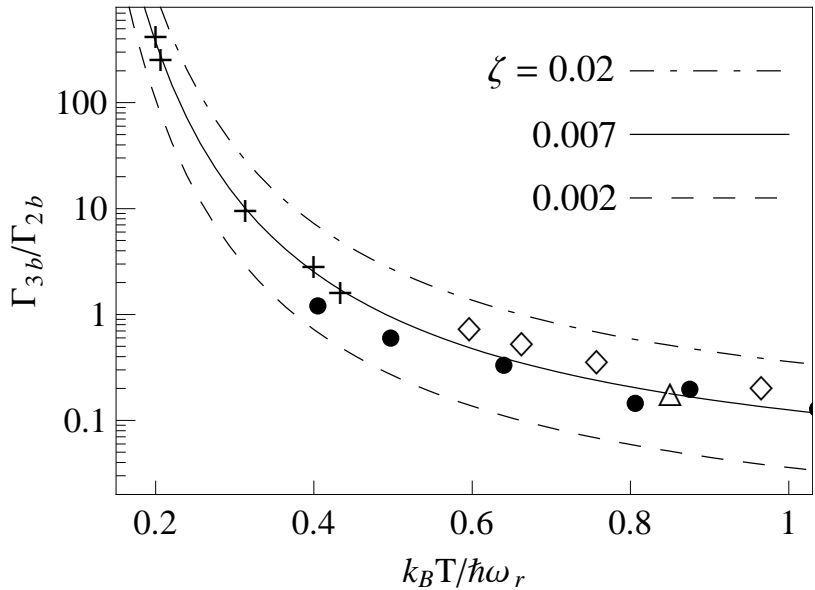,width=0.6 \columnwidth}
\caption{Ratio between the scattering rates for the two routes to thermalization  in quasi-1D systems: 
$\Gamma _{3b}$ for the effctive three-body 
collisions and $\Gamma _{2b}$ for two-body collisions leading to excited 
transverse states. The points represent the  ratios evaluated 
for various sets of experimental parameters from \cite{js1}
(points), \cite{js2} (crosses), \cite{va} (triangle), and \cite{Manz09a} (diamonds). 
In these experiments the parameter $\zeta $ was often close to 0.007 (the  ratio $\Gamma _{3b}/\Gamma _{2b}$  
for $\zeta =0.007$ exactly is shown by the solid curve). For comparison, we plot also 
$\Gamma _{3b}/\Gamma _{2b}$ for $\zeta =$ 0.002 (dashed curve)  and 0.02 (dot-dashed curve). 
Units on the axes are dimensionless. }
\label{fig:ScatRat}
\end{figure}

We can now compare the scattering rates for the two-body collisions $\Gamma_{2b}$ (Eq.~(\ref{eq:20})) or effective three-body collisions $\Gamma_{3b}$ (Eq.~(\ref{eq:17b})) and evaluate their contributions to thermalization and the breakdown of integrability in 1D systems. For $k_BT<\hbar \omega _r$ we find a simple scaling:
\begin{equation}\label{eq:ScatRat}
    \frac{\Gamma_{3b}}{\Gamma_{2b}} = \frac{C_{3b}}{C_{2b} } \; \zeta \, e^{\frac {2\hbar \omega _r}{k_BT}} = 
        \frac{3\sqrt{3}\xi ^2}{2\sqrt{2} } \; \zeta \, e^{\frac {2\hbar \omega _r}{k_BT}} \approx
        2.43 \; {\zeta}{e^{ \frac {2\hbar \omega _r}{k_BT}}}  . 
\end{equation}
The relative importance of two-body collisions and the effective three-body collions mediated by {\em virtual} excitations is determined by the dimensionless quantity $\zeta \, e^{\frac {2\hbar \omega _r}{k_BT}}$. For large $\zeta $ and small temperatures ($ k_BT \ll \hbar \omega _r $) the three-body scattering rate due to {\em virtual} excitations dominates, and can lead to thermalization  even when the thermalization processes due to two-body collisions are frozen out. For example, in typical atom chip experiment \cite{js1,va,js2} the three body reate $\Gamma_{3b}$ dominates above the two-body collisions at $ k_BT \leq \frac{1}{2} \hbar \omega _r $.  A  detailed comparison of the two rates $\Gamma_{2b}$ and $\Gamma_{3b}$ and their relation to typical experimental parameters is given in Fig. ~\ref{fig:ScatRat}.  
The scattering rate due to {\em virtual} excitations of the radial modes can dominate over real excitations for typical parameters of the recent experiment \cite{js2}.

The above calculation was for a non-degenerate ultracold gas. In a degenerate gas we need to consider the Bogoliubov-type spectrum of elementary excitations \cite{b47}, that is phononic in the long-wavelength regime, as well as the relation between atoms and elementary excitations via the Bogoliubov transformations and bosonic amplification of scattering to modes, which are initially occupied. 

Taking into account all these factors, we find the rate of damping of a fast particle in a quasicondensate (see also Ref. \cite{Glazman}): 
\begin{eqnarray}
\Gamma _{k_0}^\mathrm{damp}&=&\frac {9\sqrt{3} \xi ^2\omega _r \zeta ^2}{2\pi } 
\int _{-\infty }^\infty dk^\prime _c \int _{-\pi }^\pi d\chi ^\prime 
\int _0^\infty d\tilde{k}^\prime  \, \tilde{k}^\prime (1+n_{k_1^\prime })(1+n_{k_2^\prime })(1+n_{k_3^\prime })
\times \nonumber \\ && 
S_{k_1^\prime }S_{k_2^\prime }S_{k_3^\prime }
   \delta (k^\prime _c-k_0/\sqrt{3} )\delta (\eta _{k_1^\prime } +\eta_{k_2^\prime }+\eta _{k_3^\prime }- \eta _{k_0} ) . 
\label{eq:rates6} 
\end{eqnarray}
Here the momenta of the scattered elementary excitations are defined by the expressions reciprocal to 
Eq. (\ref{eq:rates4},~\ref{eq:rates4bis}): 
\begin{eqnarray} 
k^\prime _1&=&\frac{k_c^\prime }{\sqrt{3}} +\sqrt{\frac 23} \tilde{k}^\prime \cos (\chi ^\prime -2\pi /3), \quad  
k^\prime _2 = \frac{k_c^\prime }{\sqrt{3}} +\sqrt{\frac 23} \tilde{k}^\prime \cos (\chi ^\prime +2\pi /3), \nonumber \\  
k^\prime _3&=&\frac{k_c^\prime }{\sqrt{3}} +\sqrt{\frac 23} \tilde{k}^\prime \cos \chi ^\prime . 
\label{eq:rates7} 
\end{eqnarray} 
The energy of a mode with the momentum $\hbar k$ is $\varepsilon _k=\hbar ^2\eta _k /(2m)$ with 
$$
\eta _k=\sqrt{k^2 (k^2+8n_{1D}\alpha _s/l_r^2)}. 
$$
The static structure factor of a quasicondensate is 
$$
S_k=k^2/\eta _k.
$$
In equilibrium the population of the elementary mode with the momentum $\hbar k$ is given by the Bose-Einstein statistics with the 
mean occupation number for the mode with the momentum $\hbar k$  
$$
n_k=\frac 1{\exp [\varepsilon _k/(k_BT) ]-1} . 
$$
The initial kinetic energy $\hbar ^2k_0^2/(2m)$ of the fast atom  is assumed here to be large compared to both the mean-field interaction energy per 
particle in the quasicondensate and the temperature: 
\begin{equation}
k_0^2\gg n_{1D}\alpha _s/l_r^2, \qquad k_0^2 \gg mk_BT/\hbar ^2 
\label{eq:rates8}
\end{equation} 
Under condition (\ref{eq:rates8}) one of the scattered particles is always fast, and one of the three structure factors appearing in 
Eq. (\ref{eq:rates6}) is always very close to 1 (and the corresponding occupation number is close to 0). In the most scattering events the 
other two particles are also fast, and, hence, $S_{k^\prime _j}\approx 1$ and $n_{k^\prime _j}\approx 0$ for all three particles, $j=1,2,3$. 
Only for the scattering events with small transferred momentum two of the structure factors are significantly less than 1 and/or the 
corresponding populations approach the high-temperature limit $k_BT/\varepsilon _{k^\prime _j}$. However, in the practically interesting 
case where $k_BT\sim \hbar \omega _r n_{1D}\alpha _s$ the contribution of scattering events with small transferred momentum is  relatively small, 
and 
\begin{equation}
\Gamma _{k_0}^\mathrm{damp}\approx \frac {9\sqrt{3}}2{\xi ^2\omega _r \zeta ^2}.
\label{fast-qc}
\end{equation} 
The result of Eq. (\ref{fast-qc}) for a fast atom in a quasicondensate is also obtained by Tan, Pustilnik and Glazman \cite{Glazman}. 

\section{Calculations of the thermalization rates}
\label{sec:thermalization}

We now turn to quantify thermalization in tightly confined Bosons in a quasi-1D geometry by both two body collisions and the effective interaction (\ref{eq:11a}).  We again for simplicity consider a non-degenerate, weakly-interacting (the Lieb-Liniger parameter \cite{LL} $\gamma =2\alpha _s/(n_{1D}l_r^2)$ being much less than 1) gas of bosonic atoms. The assumptions of weak interaction and non-degeneracy enable us to express the three-particle distribution function through the product of single-particle distribution functions $f_k$. In contrast, calculation of relaxation via three-body collisions of low-energy excited states in the stronger interacting regime and especially for $\gamma \geq 1$, requires to consider the (strong) quantum correlations in the quasi-1D bosonic system. We will discuss the effects of correlations on scattering rate $\Gamma _{3b}$ and on thermalization in section \ref{sec:correlations} 

We start be writing the Boltzmann equation with a three-body collision integral \cite{beq3b}, taking into account the indistinguishability of the particles:
\begin{equation}
    \frac d{d t} f_k=\Gamma_{3b}\int _{-\infty}^\infty dk^{\prime } \int _{-\infty}^\infty dk^{\prime \prime } 
\int _{-\pi }^\pi \frac  {d\gamma }{2\pi }\left( f_{K_0 } f_{K_{-1} } f_{K_{+1} } -
  f_k f_{k^{\prime }}f_{k^{\prime \prime }} \right) , 
    \label{eq:12}  
\end{equation} 
with
\begin{eqnarray}
K_s&=&\frac {k+k^\prime +k^{\prime \prime }}3+\sqrt{\frac 23}\tilde{k}\cos (\gamma+2s\pi /3), \qquad s=0,\pm 1, 
\label{eq:120} \\
\tilde{k}&=&\sqrt{k^2+k^{\prime \, 2}+k^{\prime \prime \, 2} -\frac {(k+k^\prime +k^{\prime \prime })^2}3} . 
\label{eq:1200}
\end{eqnarray} 
Eq. (\ref{eq:12}) can be easily understood: After integration over $k^\prime $ and $k^{\prime \prime }$ the loss term in 
Eq. (\ref{eq:12}) is simply $-\Gamma _{3b}f_k$, which is the elastic three-body collision rate per atom. On the other hand, the three-atom state $| k,k^\prime ,k^{\prime \prime } \rangle $ is populated by elastic three-body collisions from those states $|K_0,K_{-1},K_{+1}\rangle $ which have the same center-of-mass momentum: $K_0+K_{-1}+K_{+1}=k+k^\prime +k^{\prime \prime }$. Since the kinetic energy of the relative motion, $\hbar ^2\tilde{k}^2/(2m)$ is conserved, the states $|K_0,K_{-1},K_{+1}\rangle $ (from where the state $| k,k^\prime ,
k^{\prime \prime } \rangle $ can be populated from) can be fully parametrized by the hyperangle $\gamma $. 

We now use the following {\em Ansatz} for the perturbed momentum distribution
\begin{eqnarray}
f_k(t)= \frac {n_{1D}}{\sqrt{\pi }k_{th}} \exp (-k^2/k_{th}^2)
[ 1+\varepsilon _4(t)H_4 (k/k_{th})] , \label{eq:16}
\end{eqnarray}
to solve Eq.~(\ref{eq:12}).  Therby $k_{th}={\sqrt{2mk_BT}/\hbar }$ and $H_4$ is the Hermite polynomial of the 4th order. This \textit{Ansatz} is the simplest nontrivial perturbation that
retains $\int dk\, kf_k=0$. We then proceed to linearizing Eq.~(\ref{eq:12}) with respect to the perturbation amplitude $\varepsilon _4(t)$ and obtain an exponential solution $\varepsilon _4(t)=\varepsilon _4(0) \exp (-\Gamma _{[4]}^{3b}t)$ with  
\begin{eqnarray}
    \Gamma _{[4]}^{3b}&=&C_{[4]}\frac {\hbar}{m} \left( \frac {n_{1D}}{l_r ^2} \right)^2 \alpha _s ^4 = C_{[4]} \, \omega_r \, \zeta^2
    \label{eq:17}
\end{eqnarray}
with the numerical constant 
\begin{equation} 
C_{[4]}=\frac 8{27}C_{3b} =\frac {8\xi ^2}{3\sqrt{3}}\approx 2.04. 
\label{eq:170} 
\end{equation} 

\begin{figure}
 \epsfig{file=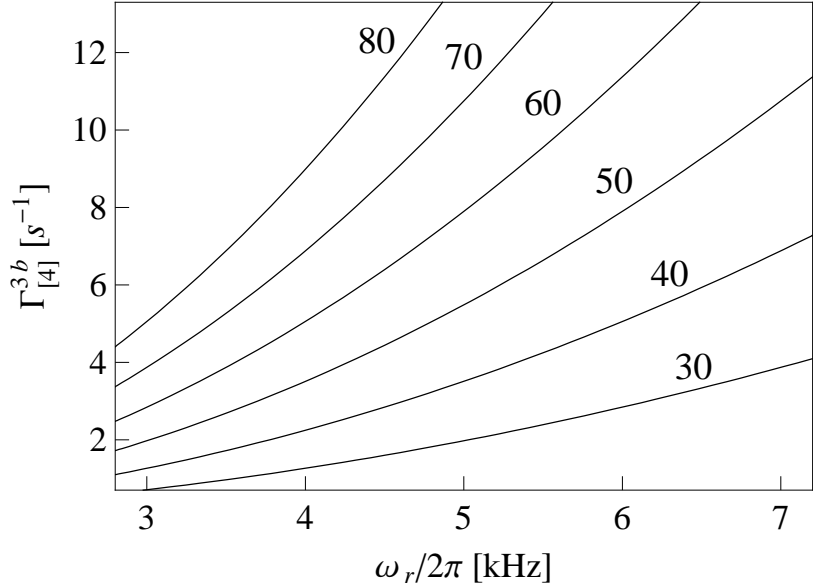,width=0.6\columnwidth}
\caption{Dependence of the rate $\Gamma _{[4]}^{3b}$ of
thermalization induced by effective three-body collisions in a weakly-interacting, quasi-1D $^{87}$Rb gas on the
radial trapping frequency for the linear densities $n_{1D} $ from $ 80~\mu $m$^{-1}$ to $30~\mu $m$^{-1}$ (from top to bottom) 
with the step $10~\mu $m$^{-1}$
(dot-dashed curve). }   \label{fig:Thermalization}
\end{figure}

To estimate the validity of our \textit{Ansatz} we note that using a higher-order Hermite polynomial $H_n$ in Eq. (\ref{eq:16}) leaving the functional dependence on the parameters of the system unchanged and leads only to a minor modification of the numerical prefactor.  For example, for $n=5$ and 6  the thermalization rates are given  by $\frac {10}{27}C_{3b} \, \omega_r \, \zeta^2$ and $\frac {34}{81}C_{3b} \, \omega_r \, \zeta^2$, respectively. It is interesting to note that in these 1D systems the thermalization rate due to the effective three-body collisions ($\Gamma _{[4]}^{3b}$) is about a factor 3 smaller then the collision rate ($\Gamma _{3b}$).  This suggests that in 1D systems thermalization requires also about 3 collisions, similar to 3D \cite{WuFoot}. Fig.~\ref{fig:Thermalization} shows numerical values of $\Gamma _{[4]}^{3b}$ as a function of the 1D density of $^{87}$Rb atoms and the radial trapping frequency. 

For comparison we calculate numerically  the thermalization rate $\Gamma _{[4]}^{2b} $ for two-body collisions involving the {\em real } transitions between the ground and excited radial states. We follow hereby the same \textit{Ansatz} and perturb the velocity distribution of atoms in the ground and excited state as given by Eq. (\ref{eq:16}), the
Boltzmannian distribution of overall populations between the levels being kept intact. In the parameter range of interest we find numerically $\Gamma _{[4]}^{2b}\approx (0.33\pm 0.03)\Gamma _{2b}$, i.e.
\begin{equation}
 \Gamma _{[4]}^{2b} \approx 0.93\,
 \omega _r \zeta e^{-\frac{2\hbar \omega _r}{k_BT} }.
 \label{eq:30}
\end{equation}
The ratio of the thermalization rates for the two-body and three-body processes is therefore very close to the respective ratio of the collision rates, shown in Fig.~\ref{fig:ScatRat}.

It is interesting to note that we find for both processes that thermalization in 1D needs about 3 collisions capable to distribute energy. This is very close to the 2.7 collisions required for thermalization in 3D \cite{WuFoot}.

For the typical parameters of an ultracold $^{87}$Rb gas on an atom chip 
\cite{js2} ($\omega _r\approx 2\pi \times 3$~kHz, $n_{1D}\approx
50~\mu $m$^{-1}$) we obtain $\Gamma _{[4]}^{3b}\approx 2$~s$^{-1}$.
This thermalization rate is temperature-independent and much
larger than the one calculated from the simple two-body collisions
with the energy sufficient to excite radial modes
$\Gamma _{[4]}^{2b}\approx 3 \times 10^{-3}$~s$^{-1}$ at the lowest
temperatures measured (30 nK).  The estimated $\Gamma _{[4]}^{3b}$
is consistent with the time needed for evaporative cooling of a
$^{87}$Rb gas on an atom chip well below $\hbar \omega _r$
\cite{js1,js2}.

\section{Suppression of thermalization by atomic correlations}
\label{sec:correlations}

The thermalization rate $\Gamma ^{3b}_{[4]} $ given by Eq. (\ref{eq:17}) was calculated for a weakly-interacting, non-degenerate gas. Calculation the thermalization rate $\Gamma ^{3b\,G}_{[4]}$ in a general case requires to take into account additional physics.  First we need to consider the effects of quantum degeneracy and  second the fact that the disperison relations for the elementary excitations in a 1D quantum (degenerate) system may differ significantly compared to a free particle, especially for phonon-like excitations. These effects, together with the bosonic amplification of the scattering into thermally populated modes, tend to accelerate thermalization. A third observation is that the three body rates require three particles to be close to the same location.  Quantum mechanically this is characterized by the third order correlation function $g_3(0)$. A full consideration of the above competing effects will require extended numerical analysis of many particular cases and transcends beyond the scope of the present manuscript. We will give here physical arguments of what to expect. 

We start by pointing out that the form of the secondary-quantized Hamiltonian Eq.~(\ref{eq:11b}) allows us to give a simple estimate of the ratio of these two rates 
$$ {\Gamma ^{3b\, G}_{[4]}}/{\Gamma ^{3b}_{[4]}} = \varrho  ({g_3(0)}/6)^2 \, . $$
Here $\varrho $ is a phase-space factor accounting for the dispersion law of elementary excitation, which changes from free-particle-like to phonon-like. But it changes the thermalization rate less dramatically than the second factor associated with the local three-body correlation function 
$$g_3(0) =\langle \hat{\tilde{\psi }}_{1D}^{\dag \,3}(z)\hat{\tilde{\psi }}_{1D}^3(z) \rangle /n_{1D}^3 \, ,$$
and will become the dominating factor when approaching the strongly correlated regime ($\gamma > 1$). For a \textit{non-degenerate} weakly-interacting Bose gas $g_3(0)=3!=6$. For a {\em degenerate} 1D Bose gas $g_3(0)$ has been recently calculated for the whole range of atomic repulsion strength ($0<\gamma <\infty $) by Cheianov et al. \cite{G3gen}. In the zero-temperature limit $g_3(0)$ rapidly decreases from 1 to $16\pi ^6/(15\gamma ^6)$ as $\gamma$ grows from 0 to values $\gamma \gg 1$.

We now turn to the above conjecture on suppression of the thermalization by atomic correlations. A detailed calculation can be found in Ref. \cite{we2}, here we sketch the basic physics argument.  To look at the correlations we start by considering $N$ identical bosons in 1D configuration with the Hamiltonian \ref{eq:11b}, which, after rescaling of units, takes the form 
\begin{eqnarray}
\hat{H}&=&-\sum _{j=1}^N \frac {\partial ^2}{\partial z_j^2} + 2c
\sum _{j>j^\prime } \delta (z_j-z_{j^\prime }) +  
\sum _{j>j^\prime >j^{\prime \prime }}U_\mathrm{3b}
(z_j-z_{j^\prime },\, z_j-z_{j^{\prime \prime }}). \label{eq1}
\end{eqnarray}
$c=2\alpha _s/l_r ^2$ is the strength of interaction of two atoms in the tight waveguide with ground state size $l_r$.  $U_\mathrm{3b}$ is obtained by by adiabatic elimination of transverse modes virtually excited by the 3D short-range pairwise atomic interaction \cite{we1}.  
The explicit form of  $U_\mathrm{3b}$ is given by Eq. (\ref{U3bIIIA}) within a numerical prefactor, 
$U_\mathrm{3b}=[\hbar ^2/(2m)]{\cal U}_\mathrm{3b}$.

We follow now our detailed calcultions in Ref. \cite{we2} and estimate the three-body scattering amplitude in the presence of the delta-functional pairwise repulsive interactions. The stronger the {\em pairwise} interparticle repulsion, the smaller is the probability of a close encounter of three particles which will result in a suppression of the three-body scattering amplitude. The simplest case is to analyze the Hamiltonian (\ref{eq1}) for $N=3$ particles.  For that purpose we express it in hyperspherical coordinates $R$, $\chi $ defined as \cite{o1}
\begin{equation}
Z_c=\frac {z_1+z_2+z_3}{\sqrt{3}} ,\quad 
R \sin \chi =\frac {z_1-z_2}{\sqrt{2}} , \quad
R \cos \chi =\sqrt{\frac 23}\left( z_3-\frac {z_1+z_2}2\right)
\label{eq2}
\end{equation}
and obtain  for the Hamiltonian 
\begin{eqnarray}
\hat{H}&=&-\frac {\partial ^2}{\partial Z_c^2} -\frac 1R
\frac \partial {\partial R}R\frac \partial {\partial R}
-\frac 1{R^2}\frac {\partial ^2}{\partial \chi ^2} + 
\frac {\sqrt{2}c}R\sum _{\nu =-2}^3 \delta (\chi -\nu \pi /3) +
U_\mathrm{3b} (R,\chi ).
\label{eq3}
\end{eqnarray} 
The corresponding Schr\"odinger equation for the three-particle wave function is
$$\hat{H}\Psi (z_1,z_2,z_3) =(k_1^2+k_2^2+k_3^2)\Psi (z_1,z_2,z_3) \, .$$ 
The wavenumbers $k_j$ are defined from the set of transcendental equations \cite{LL}, provided that the periodic boundary conditions are set on the interval of the length $L$. By setting $L\rightarrow \infty $, we obtain a continuous spectrum, where $k_j$'s are real for repulsive interaction ($c>0$). We can now separate the center-of-mass motion and describe the relative motion in hyperspherical coordinates.  This leads to the \textit{Ansatz}: 
$$\Psi (z_1,z_2,z_3)=\exp [i(k_1+k_2+k_3)(z_1+z_2+z_3)/3] \psi _\mathrm{r}(R,\chi )$$
where the kinetic energy of the relative motion is given by 
$$k^2=\frac 13 [(k_1-k_2)^2+(k_2-k_3)^2+(k_3-k_1)^2]$$. 
In the adiabatic hyperspherical approximation \cite{mc1}, which holds in the long-wavelength limit 
\begin{equation} 
k\ll c    \label{eq.lwll}
\end{equation}   
and neglects coupling of different scattering channels as well as accumulation of phase shifts of the scattered wave due to non-adiabatic effects we get 
\begin{equation}
\psi _\mathrm{r}(R,\chi )= F_0(R)B _0(R,\chi ). 
\label{eq60}
\end{equation}
The hyperangular part $B _0(R,\chi )$ of this wave function is the eigenfunction of the Hamiltonian (\ref{eq3}) with fixed $R$, corresponding to the lowest eigenvalue $\lambda _0(R)$, which is the smallest positive root of the transcendental equation 
\begin{equation}
\lambda (R) \tan [ \pi \lambda (R)/6]=cR/\sqrt{2}  .
\label{eq0110}
\end{equation}
Using the regular hexagon symmetry group \cite{d6} of the Hamiltonian (\ref{eq3}), we write the hyperangular part of Eq. (\ref{eq60}) as 
\begin{equation}
B _0(R,\chi )=\tilde{B}_0(R,\chi )+\tilde{B}_0(R,\chi -2\pi /3)+
\tilde{B}_0(R,\chi +2\pi /3),
\label{eq90}
\end{equation}
where 
\begin{equation}
\tilde{B}_0(R,\chi )=\left \{ \begin{array}{ll} \cos [\lambda _0(R)(\pi /6-|\chi |)] , & | \chi |\leq \pi /3 \\
0 ,  & \mathrm{otherwise} \end{array} \right.
\label{eq103}
\end{equation}
After integrating out the hyperangular variable, the Schr\"odinger equation in the adiabatic hyperspherical approximation reduces to 
\begin{equation}
-\frac 1R \frac d{dR}R\frac d{dR} F_0+\left[ \frac {\lambda _0^2(R)}{R^2}+
\tilde{U}_{00}(R)\right] F_0=k^2F_0
\label{eq17}
\end{equation}
where 
\begin{equation}
\tilde{U}_{00 }(R)=\frac { \int _0^{\pi /3}d\chi \,
{\tilde{B}_{0}^2(R,\chi )U_\mathrm{3b}(R,\chi )}}{  \int _0^{\pi /3}d\chi \,
{\tilde{B}_{0}^2(R,\chi )}} .
\label{eq201}
\end{equation} 
with the boundary conditions requiring $F_0$ to be finite for both $R=0$ and $R\rightarrow \infty $, for the ``partial wave" corresponding to the lowest eigenvalue $\lambda _0(R)$, whose
asymptotic expressions are
\begin{equation}
\lambda _0(R)\approx \left \{ \begin{array}{ll}
\sqrt{ \frac {3\sqrt{2}cR}\pi },&cR\ll 1\\
3-\frac {18\sqrt{2}}{\pi cR}, &cR\gg 1\end{array}\right.  .
\label{eq18}
\end{equation}
We can solving now  Eq. (\ref{eq17}) analytically in two regions, $cR\ll 1$ and $cR\gg 1$, with $\lambda _0(R)$ approximated by Eq. (\ref{eq18}), and tailoring the solutions by quasiclassical expressions for $F_0(R)$ in the intermediate hyperradius range.  The scattering amplitude $\tilde{f}_0$ (for its definition in planar geometry see \cite{shl2da,ldrs,ajp81,ajp86}) can then be obtained from the asymptotic form  of the wave function at $R\rightarrow \infty $  
\begin{equation}
F_0(R)\approx  J_3(kR)-i\tilde{f}_0H^{(1)}_3(kR) ,   
\label{eq28}
\end{equation}
where $H^{(1)}_3(z)=J_3(z)+iY_3(z)$ is the Hankel function of the
first kind and $J_3(z),~Y_3(z)$ are the third-order Bessel functions. 
The behavior of $F_0$ at $R\rightarrow 0$ is defined by the details of the potential $U_{00}(R)$, but the result can be 
finally expressed via the effective vertex of the three-body elastic collisions, thus yielding 
\begin{equation}
\tilde{f}_0=\frac {6(\alpha _s/l_r)^2\xi }{2\pi \Omega  \left( \frac ck\right) ^6 -i \pi  } , 
\label{eq209}
\end{equation}
where $\Omega  \approx 1$ is a numerical constant and $\xi \approx 1.15$ is defined by Eq. (\ref{defxi}). We use thereby the fact that $U_{00}(R)$ differs significantly from zero on the length scale $l_r$, over which a virtually excited particle can propagate, and $c l_r\ll 1$.

From  Eq. (\ref{eq209}) we conclude that the three-body scattering amplitude decreases in proportion to $(k/c)^6$ as $k/c \rightarrow 0$, i.e., when the interaction is strong enough to induce significant atomic correlations. The three-body scattering rate in a 1D system of bosons in the case of strong pairwise interaction is suppressed by a factor $\sim (k/c)^{12}$. By averaging over collision momenta in a moderately-excited strongly-interacting state we obtain the scattering rate suppression factor $\sim \langle (k/c)^{12}\rangle \sim \gamma ^{-12}$.

We can now compare our result with the zero-distance three-particle correlation function $g_3(0)$. In the strong interaction limit $\gamma \gg 1$   $g_3(0)\propto \gamma ^{-6}$  \cite{G3gen,g3a} and gives a direct physical motivation of our original conjecture \cite{we1} that the pairwise interactions and the quantum correlations induced by them in a strongly-interacting 1D bosonic system suppress the three-body elastic scattering rate, and, hence, thermalization, by a factor $\propto g_3^2(0)$. In other words, strong quantum correlations extend the time scale, 
on which a quasi-1D system approaching the Tonks-Girardeau regime can be considered as approximately integrable. 

Moreover, we can corroborate this conjecture by observing that the thermalization rate is proportional to the square of the matrix element of the transition operator that is proportional to $\int dz\, \hat{\tilde{\psi }}^\dag \hat{\tilde{\psi }}^\dag \hat{\tilde{\psi }}^\dag \hat{\tilde{\psi }} \hat{\tilde{\psi }} \hat{\tilde{\psi }}$. Since the energies of the products of the {\em elastic } three-body process are low (of about $k_BT$), we may assume that the correlations in the initial and the final states are the same, and the transition matrix element can be regarded as proportional to $g_3(0)$ that yields again the $ g_3^2(0)$ scaling of the rate. In contrast to this, the {\em inelastic} three-body processes are accompanied by a large energy release, and after an inelastic collision the newly formed dimer molecule and the fast atom leave the system almost immediately. Therefore the inelastic three-body relaxation rate in a 1D ultracold Bose gas is proportional to the first power of $g_3(0)$ \cite{Tolra}. 

We now compare the calculated thermalization rates to the quantum Newton's cradle experiment \cite{dw1} where a \textit{lower boundary} for the damping time of the 1D motion towards a Gaussian profile was estimated. For three different Lieb-Liniger parameters $\gamma = 1.4$, 3.2, 18 Kinoshita et al estimate lower bounds to the thermalization time from the consistency of the the experimental momentum distributions with the experimentally observed heating during the time interval of 0.5, 0.5, 1.0 seconds probed. They find one sigma lower limits of 2.6 s, 25 s, and 13 s respectively. In their experiment the motion of two groups of $^{87}$Rb atoms was excited at the relative velocity equal to 4 recoil velocities, which is far above the width of the ground-state velocity distribution.  We therefore can not expect the collision rate to be suppressed in proportion to $g_3^2(0)$ that, as described above for slow collisions. Instead, we have to apply the estimates for damping of a fast particle discussed in the end of section IV.  In a strongly-correlated system one has to take the particle correlations into account and eq. \ref{fast-qc} has to be multiplied by a factor $g_2(0)$ denoting the {\em two-particle} correlation function at zero distance (Tan, Pustilnik and Glazman \cite{Glazman}):  
  \begin{equation}
  \Gamma _{k_0}^\mathrm{damp}\approx \frac {9\sqrt{3}}2{\xi ^2\omega _r \zeta ^2g_2(0)}.
  \label{fast-g}
  \end{equation} 
Substituting the experimental parameters of Ref. \cite{dw1} and taking the values for $g_2(0)$ from Ref. \cite{gtwo}, we obtain $\Gamma _{k_0}^\mathrm{damp} \approx 15~s^{-1}$, $1.7~s^{-1}$, and $7 \times 10^{-3}~s^{-1}$ for $\gamma = 1.4$, 3.2, and 18, respectively.  To compare this calculated damping rates to the experiment in \cite{dw1} one has to consider that (1) the two colliding clouds overlap only for a very short time during each oscillation and (2) that the thermalization rate is a factor 3 longer (section V). Taking this into account we estimate the respective thermalization times of 2.6 s, 35 s and $>$ 1000 s. These rates are consistent with the experimental findings of Ref. \cite{dw1}. For a more detailed comparison one would need longer time scale experiments with lower intrinsic heating and more detailed calculations of the dynamics including the damping due to three body collisions discussed here and in \cite{Glazman}.

\section{Conclusion} 

A radially confined atomic gas is never perfectly 1D, and radial motion can be excited, either in reality or virtually even if Eq. (\ref{eq:1}) holds. This possibility leads to effective three-body collisions, which arise in the second order of perturbation theory and can be associated with virtual excitation of radial modes. These processes lead to thermalization even when two body collisions are frozen out at $k_BT \ll \hbar \omega_r$ and provide a mechanism to break integrability in 1D systems. In other words, the freeze-out of the radial modes is only a \textit{necessary}, but not sufficient condition for integrability in 1D systems.  Our estimations of the relaxation rates for weakly interacting quasi-1D Bose gases are consistent with recent experimental observations for weakly-interacting quasicondensates \cite{js1,js2}.

These effective three-body collisions can be suppressed by quantum correlations caused by strong pairwise repulsions. If they dominate, as in a strongly correlated 1D Tonks-Girardeau gas, they suppress the influence of the integrability-breaking interaction term. The thermalization rate decreases in proportion to $g_3^2(0)$ as the system enters the regime of strong correlations ($\gamma \gg 1$), and the system (remaining non-integrable in the strict sense) behaves like (almost) integrable on time scales short compared to the inverse thermalization rate.

The effective three-body collisions, and their suppression by quantum correlations should be accessible in experiments looking at the damping of fast, particle-like excitations in systems with $\gamma $ varying in a broad range of values from less than $1$ to $\sim 10$.

This work is supported by the EC (STREP MIDAS) and the FWF.

\end{document}